\documentclass[12pt]{iopart}
\usepackage{amssymb}
\usepackage{graphicx}
\usepackage{subfigure}
\usepackage{epstopdf}
\usepackage{color}
\usepackage[misc]{ifsym}
\usepackage{threeparttable}
\usepackage{xfrac}
\usepackage{algorithm}
\usepackage{algorithmicx}
\usepackage[noend]{algpseudocode}
\usepackage{dsfont}
\usepackage{bm}
\usepackage{subfigure}
\usepackage{booktabs}
\usepackage{multirow}

\newcommand{\ie}{\emph{i.e., }}
\newcommand{\eg}{\emph{e.g., }}
\newcommand{\aka}{\emph{a.k.a.,}}

\begin{document}

\title[Quantum Architecture Search with Meta-learning]{Quantum Architecture Search with Meta-learning}

\author{Zhimin He$^{1\dag}$, Chuangtao Chen$^{2\dag}$, Lvzhou Li$^{3}$, Shenggen Zheng$^4$, Haozhen Situ$^{5*}$}

\address{$^1$School of Electronic and Information Engineering, Foshan University, Foshan 528000, China}
\address{$^2$School of Mechatronic Engineering and Automation, Foshan University, Foshan 528000, China}
\address{$^3$Institute of Quantum Computing and Computer Science Theory, School of Computer Science and Engineering, Sun Yat-Sen University, Guangzhou 510006, China}
\address{$^4$Peng Cheng Laboratory, Shenzhen 518055, China}
\address{$^5$College of Mathematics and Informatics, South China Agricultural University, Guangzhou 510642, China}
\ead{situhaozhen@gmail.com}
\vspace{10pt}
\begin{indented}
\item[$ \dag $: The first two authors contribute equally to this work. ]
\end{indented}

\begin{abstract}
Variational quantum algorithms (VQAs) have been successfully applied to quantum approximate optimization algorithms, variational quantum compiling and quantum machine learning models. The performances of VQAs largely depend on the architecture of parameterized quantum circuits (PQCs). Quantum architecture search (QAS) aims to automate the design of PQCs in different VQAs with classical optimization algorithms. However, current QAS algorithms do not use prior experiences and search the quantum architecture from scratch for each new task, which is inefficient and time consuming. In this paper, we propose a meta quantum architecture search (MetaQAS) algorithm, which learns good initialization heuristics of the architecture (\ie meta-architecture), along with the meta-parameters of quantum gates from a number of training tasks such that they can adapt to new tasks with a small number of gradient updates, which leads to fast learning on new tasks. The proposed MetaQAS can be used with arbitrary gradient-based QAS algorithms. Simulation results of variational quantum compiling on three- and four-qubit circuits show that the architectures optimized by MetaQAS converge much faster than a state-of-the-art gradient-based QAS algorithm, namely DQAS. MetaQAS also achieves a better solution than DQAS after fine-tuning of gate parameters.
\end{abstract}

\section{Introduction}
Current available quantum devices are noisy intermediate-scale quantum\,(NISQ) devices\,\cite{PJ18}. In NISQ era, variational quantum algorithms (VQAs) make great success due to their resilience to errors and high flexibility with respect to quantum resource requirements including circuit depth and numbers of qubits \cite{KMT17,higgott2019variational,jones2019variational,farhi2014quantum,moussa2020quantum,cerezo2021variational}. VQA minimizes a cost function calculated by a parameterized quantum circuit (PQC). The cost function is evaluated on quantum devices while the optimization of gate parameters in PQC occurs on classical computers.  VQAs have been successfully applied to ground and excited state preparations\,\cite{KMT17,higgott2019variational,jones2019variational}, variational quantum compiling\,\cite{KLP19, zhang20,he2021variational} and approaching better approximation for combinatorial optimization problems\,\cite{farhi2014quantum,moussa2020quantum,hadfield2019quantum,morales2020universality}.
A large number of quantum machine learning algorithms based on PQC have been proposed, including quantum neural networks\,\cite{li2020quantumCNN,wei2021quantum}, quantum generative models\,\cite{liu2018differentiable,SHW20,hu2019quantum} and clustering algorithm\,\cite{OMA17}.
These algorithms are not restricted to theory, and have been tested on real devices such as IBM's and Rigetti's quantum devices.

The aforementioned VQAs are realized by PQCs. In typical VQAs, the architectures of PQCs are fixed and only the parameters of quantum gates are trained to optimize an objective function, which is inflexible and undesirable. The performances of such VQAs largely depend on the architectures of PQCs, as the PQCs with different architectures may differ greatly in their expressive and entanglement capacities\,\cite{akshay2020}.
Most of current VQAs manually design quantum circuits, which rely on human expertise and require hundreds of experimentations or modifications from the existing quantum circuits.

Automatic design of quantum circuits has become a hot topic in VQAs, which does not require human expertise and can find out quantum circuits with equivalent or better performance than manually designed ones \cite{zhang20,he2021variational,rattew2019domain,zhang2021neural,zhang2020differentiable,kuo2021quantum,li2020quantum, ostaszewski2021structure, grimsley2019adaptive,sharma20,cincio20,chivilikhin2020mog,ostaszewski2021reinforcement}.
Zhang \etal introduced the concept of quantum architecture search (QAS) \cite{zhang2021neural,zhang2020differentiable} to represent a collection of methods that automatically search for an optimal quantum circuit for a given VQA.
Automatic design of quantum circuits is also denoted as ansatz architecture search \cite{li2020quantum}, structure optimization for PQCs \cite{ostaszewski2021structure}, adaptive variational algorithm \cite{grimsley2019adaptive}, evolutional VQE \cite{rattew2019domain} and circuit learning \cite{cincio20} in different VQAs.
We use the term QAS in the paper as it covers all scenarios of quantum circuit design.

QAS is treated as a discrete optimization problem, which can be solved by gradient-free optimizers \eg simulated annealing \,\cite{sharma20,cincio20}, evolutionary algorithms\,\cite{rattew2019domain,chivilikhin2020mog} and reinforcement learning\,\cite{zhang20,he2021variational,kuo2021quantum,ostaszewski2021reinforcement}.
However, these algorithms are time consuming and require a large amount of computing resources as they require to evaluate a large amount of quantum circuits.
Accelerating the searching process of QAS algorithms is crucial, especially for the applications which require to generate quantum circuits within limited time.
In order to accelerate the searching process, weight sharing strategy is used in one-shot QAS, where the trainable gate parameters are reused instead of optimizing them individually for each ansatz\,\cite{du2020quantum}. In order to speed up the evaluations of quantum architectures, a neural predictor is trained to directly predict the performance based on the architecture alone rather than training a quantum circuit and then evaluating its performance\,\cite{zhang2021neural}.
Compared to gradient-free QAS, gradient-based QAS algorithms are orders of magnitude faster to get the optimal quantum circuit by gradient descent.
Zhang \etal proposed a gradient-based QAS algorithm by transferring the discrete architecture search problem into a continuous optimization problem and optimizing  the quantum architecture by gradient descent\,\cite{zhang2020differentiable}.

The aforementioned  QAS algorithms do not use prior experiences and search the quantum architecture from scratch for each new task, which is inefficient and time consuming.
Another strategy to speed up the searching process is finding good initialization heuristics (\ie prior experiences) of architecture parameters, which enable rapid convergence to local minima of the loss function.
Previous researches have successfully found good initialization heuristics of gate parameters with meta-learning\,\cite{verdon2019learning,wilson2021optimizing}.
The numerical experiments on quantum approximate optimization algorithms and variational quantum eigensolver show that the initialization heuristics of gate parameters learned by meta-learning enable the cost function to converge faster. Besides, these algorithms are more resistant to noise. However, these researches only focused on the initialization heuristics of gate parameters, without any consideration of the architecture of PQCs.

In this paper, we propose to learn a good initialization heuristic on the architecture of PQCs with meta-learning. Once a meta-architecture is trained from a number of training tasks, it can be adapted to different new tasks with just a few steps of the optimizer, which is much faster than current gradient-based QAS algorithms. Since the meta-architecture is trained once and the training can be offline, the training complexity is not a critical problem.
Compared with prior researches\,\cite{verdon2019learning,wilson2021optimizing}, we not only find a good initialization heuristic on gate parameters, but also learn a meta-architecture which can fast adapt to new tasks. The meta-architecture is trained based on the model-agnostic meta-learning algorithm, which does not introduce any learned parameters compared to the meta-learned neural optimizer in refs.\,\cite{verdon2019learning,wilson2021optimizing}.
Moreover, MetaQAS can be used with arbitrary gradient-based QAS algorithms. In our work, we choose a state-of-the-art gradient-based QAS algorithm, \ie DQAS \,\cite{zhang2020differentiable}.
We evaluate the performance of the proposed method on a typical kind of VQAs, \ie variational quantum compiling.
The simulation results of variational quantum compiling on three- and four-qubit circuits show that QAS algorithm starting from the meta-architecture and meta-parameters obtained by the proposed MetaQAS converges much faster and generates a better quantum circuit for VQA than the one without initialization heuristics.

The rest of the paper is organized as follows. Section 2 introduces some related work, including meta-learning, variational quantum compiling and quantum architecture search. The proposed quantum architecture search algorithm with meta-learning (MetaQAS) is described in Section 3. The performance of MetaQAS is compared to the state-of-the-art QAS algorithm in terms of the final loss and the number of iterations to converge through numerical simulation in Section 4. Finally, we summarize the results of this paper and discuss the future work in Section 5.

\section{Related work}
\subsection{Meta-learning}

When humans learn new skills, we rarely start from scratch. Instead, we begin with learned skills on early related tasks and learn new skills or tasks much faster with the help of experiences\,\cite{lake2017building}. Likewise, meta-learning,
\aka learning to learn, is the science of systematically learning experiences from a wide range of learning tasks, and then learns new tasks in a faster and more efficient way.
It provides a possible method to tackle the conventional challenges of machine learning algorithms, including computation and data bottlenecks.
Meta-learning has seen a dramatic rise in interest and achieved great successes in image recognition\,\cite{yao2019automated,flennerhag2019meta}, recommendation systems\,\cite{vartak2017meta,lu2020meta}, speech recognition\,\cite{winata2020learning,hsu2020meta} and neural architecture search\,\cite{lian2019towards,wang2020m,elsken2020meta}.


Meta-learning models are broadly divided into three categories, \ie  metric-, model- and optimization-based meta-learning.
Metric-based methods learn a good feature space to measure similarity and then use that feature space for a variety of new tasks\,\cite{sung2018learning,vinyals2016matching}, which is conceptually simple and efficient as it does not need task-specific adjustments. However, comparisons between tasks become computationally expensive when the number of tasks is large.
Model-based methods aim to design a model that can quickly update its parameters within very few training steps\,\cite{santoro2016meta,wang2016learning}. They are flexible and have broad applicability.
Optimization-based methods search for parameters that facilitate rapid gradient-based adaptation to new tasks\,\cite{andrychowicz2016learning,finn2017model,nichol2018first}. In this paper, we use optimization-based meta-learning as it is more suitable for extrapolation beyond the distribution of tasks in the meta-training set.

A typical optimization-based method is model-agnostic meta-learning (MAML)\cite{finn2017model}, which explicitly optimizes for fast adaptation to new tasks by learning proper initialization parameters.
MAML aims to search for initialization parameters $\bm \theta_{meta}$ of a machine learning model which satisfy
\begin{eqnarray}
	\bm \theta_{meta} = \arg\min_{\bm \theta}\sum_{\mathcal{T}_i\sim p(\mathcal{T})}L_{\mathcal{T}_i}(\Phi^k(\bm \theta,\mathcal{T}_i)),
	\label{eq:maml}
\end{eqnarray}
where $p(\mathcal{T})$ is a probability distribution over tasks that we want our model to be able to adapt to.
$\Phi^k(\bm \theta,\mathcal{T}_i)$ denotes the learning model at the $k$th iteration of updates from the initialization parameters $\bm \theta$ trained on task $\mathcal{T}_i$.
The meta-objective function $\sum_{\mathcal{T}_i\sim p(\mathcal{T})}L_{\mathcal{T}_i}(\Phi^k(\bm \theta,\mathcal{T}_i))$ measures the quality of the initialization parameters in terms of the total loss across all tasks.
The meta-objective function is minimized by optimizing the initialization parameters $\bm \theta_{meta}$, which contain the across-task knowledge.

The optimization of the parameters in the learning model on  task $\mathcal{T}_i$ is performed via  gradient descent.
The updated parameters of the learning model after $k$ iterations on task $\mathcal{T}_i$ can be expressed as
\begin{eqnarray}
	\bm \theta_i^k = \bm \theta_i^{k-1} - \lambda_{task} \nabla _{\bm \theta} L_{\mathcal{T}_i}(\Phi(\bm \theta_i^{k-1},\mathcal{T}_i)),
	\label{eq:mamlTask}
\end{eqnarray}
where $\lambda_{task}$ is the learning rate of the parameters for a given task. The dependence of $\bm \theta_i^k$ on the initialization parameters $\bm \theta$ is explicitly denoted by unrolling Equation\,(\ref{eq:mamlTask}).
The update for the meta initialization parameters $\bm \theta_{meta}$ is
\begin{eqnarray}
	\bm \theta_{meta}^{j+1} = \bm \theta_{meta}^{j} - \lambda_{meta}\nabla _{\bm \theta} \sum_{\mathcal{T}_i\sim p(\mathcal{T})}L_{\mathcal{T}_i}(\Phi^k(\bm \theta_{meta}^{j},\mathcal{T}_i)),
	\label{eq:maml}
\end{eqnarray}
where $\lambda_{meta}$ is the learning rate of the meta-parameters.
REPTILE\,\cite{nichol2018first}, a variant of MAML, executes stochastic gradient descent with a first-order derivative to achieve higher efficiency in computation
\begin{eqnarray}
	\bm \theta_{meta}^{j+1} = \bm \theta_{meta}^{j} + \lambda_{meta} \sum_{\mathcal{T}_i\sim p(\mathcal{T})}(\bm \theta_{\mathcal{T}_i}^k - \bm \theta_{meta}^j),
	\label{eq:reptile}
\end{eqnarray}
where $\bm \theta_{\mathcal{T}_i}^k$ is the parameters at the $k$th iteration of updates trained on task $\mathcal{T}_i$.



\subsection{Variational quantum compiling}
\begin{figure}
	\centering
	\includegraphics[width=7.9cm]{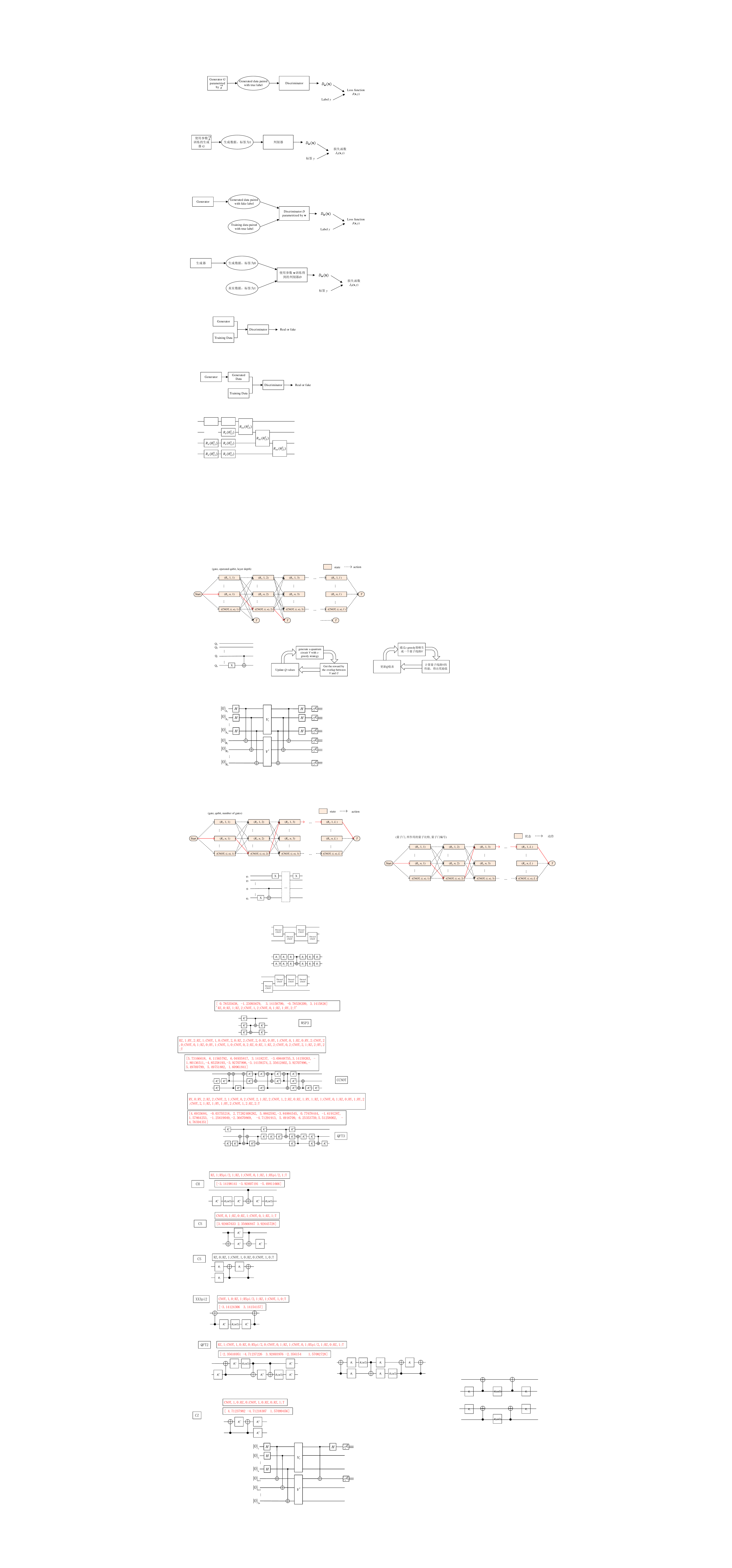}
	\caption{The quantum circuit to calculate $C_{local}^{(1)}(V_t,V(\bm l,\bm{\theta}))$.}
	\label{fig:HSTCircuit}
\end{figure}

Variational quantum compiling (VQC) compiles a target unitary $V_t$ to a trainable quantum gate sequence $V$ which has approximately the same action as $V_t$\,\cite{KLP19,sharma20,heya2018}.
The gate sequence $V$ is constructed by the native gate alphabets implemented by a specific quantum hardware, which can be denoted as $\mathcal{A} = \{G_l(\theta)\}_l$, where gate $G_l(\theta)$ is native to the quantum hardware.
The type of gate and the corresponding qubit it operates on are determined by a discrete parameter $l$.  $\theta$ denotes the parameter of the quantum gate.
Variational quantum compiling aims to search for a gate sequence $\bm l$  as well as the trainable gate parameters $\bm\theta$ to minimize the difference between
the target unitary $V_t$ and $V(\bm l, \bm\theta)$:
\begin{eqnarray}
	(\bm{l}^{*},\bm{\theta}^{*})  = \arg \min_{\bm{l},\bm{\theta}} C(V_t,V(\bm l,\bm{\theta})),
	\label{eq:obj}
\end{eqnarray}
where $V(\bm l,\bm\theta) = G_{l_N}(\theta_N)G_{l_{N-1}}(\theta_{N-1})...G_{l_1}(\theta_1)$.
$\bm l=(l_1,l_2,...,l_N)$ describes the $N$ gates selected from the native gate set and the qubits they operate on.
$\bm \theta=(\theta_1,\theta_2,...,\theta_N)$ denotes the continuous parameters associated with the selected gates.
$C(V_t,V(\bm l,\bm{\theta}))$ measures the difference between $V(\bm l,\bm{\theta})$  and $V_t$, which can be calculated by Hilbert-Schmidt Test (HST)
\begin{eqnarray}
	C(V_t,V(\bm l,\bm{\theta})) = 1-\frac{|\langle V(\bm l,\bm{\theta}), V_t \rangle|^2}{d^2}=1-\frac{|\text{Tr}( V(\bm l,\bm{\theta})^\dagger V_t)|^2}{d^2},
	\label{eq:hst}
\end{eqnarray}
where $d$ is the Hilbert-space dimension. $C(V_t,V(\bm l,\bm{\theta})) = 0$ if and only if $V_t=V(\bm l,\bm{\theta})$. However, global cost functions are easy to cause barren plateau phenomena, even for shallow circuits\,\cite{mcclean2018barren,cerezo2020cost}.
Local Hilbert-Schmidt Test (LHST)\,\cite{KLP19} is proposed to mitigate the barren plateau phenomena, which can be calculated by
\begin{eqnarray}
	C_{local}(V_t,V(\bm l,\bm{\theta})) = \frac{1}{n}\sum_{i=1}^{n}C_{local}^{(i)}(V_t,V(\bm l,\bm{\theta})),
	\label{eq:lhst}
\end{eqnarray}
where $C_{local}^{(i)}(V_t,V(\bm l,\bm{\theta}))$ is equal to the probability of the all-zeros outcome on qubits $i$ and $i+n$  for the Bell-basis measurement, which can be calculated by the quantum circuit in \textbf{Figure\,\ref{fig:HSTCircuit}}\,\cite{KLP19}.
Khatri \etal showed that LHST scaled well with problem size. $\bm \theta$ are continuous parameters which can be optimized by gradient descent. However, the gate sequence $\bm l$ are discrete, which can not be directly optimized via gradient descent.

Many quantum compiling algorithms\,\cite{sharma20,jones2018,singh19,xu19} use a specific template instead of optimizing the gate sequence $\bm l$. However, the template is usually not optimal and consists of redundant quantum gates, which increase the errors of quantum algorithms due to decoherence process and gate noise in NISQ devices.
Recently some intelligent algorithms including simulated annealing\,\cite{KLP19} and reinforcement learning\,\cite{he2021variational} are used to optimize the gate sequence $\bm l$ in variational quantum compiling.

\subsection{Quantum architecture search}
Variational Quantum Algorithms (VQAs) have become the key strategy to obtain quantum advantage on NISQ devices \cite{KMT17,higgott2019variational,jones2019variational,farhi2014quantum,moussa2020quantum,cerezo2021variational,KLP19, zhang20,he2021variational}.
VQA requires a parametrized quantum circuit (PQC) whose trainable parameters are iteratively adjusted via a classical optimizer in order to minimize an objective function.
The performance of VQA largely depends on the design of PQC. A badly designed PQC has limited expressive power and is vulnerable to barren plateau problem \cite{mcclean2018barren}.  Most of current VQAs use fixed architectures designed based on personal experience, resulting in the requirement of human expertise\,\cite{KLP19,sharma20,jones2018,singh19}.
Automatic design of quantum circuits has become a hot topic of VQAs. Zhang \etal introduced the concept of quantum architecture search (QAS) \cite{zhang2021neural,zhang2020differentiable} to represent a collection of methods that automatically search for an optimal quantum circuit for a given VQA algorithm \cite{zhang20,he2021variational,rattew2019domain,zhang2021neural,zhang2020differentiable,kuo2021quantum,li2020quantum, ostaszewski2021structure, grimsley2019adaptive,sharma20,cincio20,chivilikhin2020mog,ostaszewski2021reinforcement}.
QAS can be regarded as a discrete optimization problem, which can be solved by gradient-free optimizers
\eg simulated annealing \,\cite{sharma20,cincio20}, evolutionary algorithms\,\cite{chivilikhin2020mog,rattew2019domain} and reinforcement learning\,\cite{zhang20,he2021variational,kuo2021quantum,ostaszewski2021reinforcement}.
Current researches show that VQAs based on QAS algorithms can outperform VQAs based on human expertise in various applications.
However, these QAS algorithms search for an architecture in discrete domain and need to evaluate the performances of a large amount of quantum circuits with different architectures, which is time consuming.

Zhang \etal\,\cite{zhang2020differentiable} proposed a differentiable quantum architecture search (DQAS) to relax the discrete space of QAS onto a continuous domain and optimized the gate sequence $\bm l$ via gradient descent.
Discrete gate sequences $\bm l$ are sampled from a probabilistic model $P(\bm l,\bm \alpha)$  characterized by continuous parameters $\bm \alpha$.
The objective of DQAS is to minimize
\begin{eqnarray}
	\mathcal{L}=\sum_{\bm l \sim P(\bm l,\bm \alpha)}L(V(\bm l, \bm \theta)),
	\label{eq:DQAS}
\end{eqnarray}
where $V(\bm l, \bm \theta)$ denotes the quantum circuit determined by the architecture vector $\bm l$ and gate parameters $\bm\theta$. $L$ is the loss function of VQA algorithm.
The objective function $\mathcal{L}$ of DQAS indirectly depends on continuous parameters $\bm \theta$ and $\bm \alpha$, which can be optimized via gradient descent.
The gradient with respect to gate parameters $\bm \theta$ can be calculated with analytical gradient measurements \cite{harrow2021low} or parameter-shift rule \cite{crooks2019gradients}.
$\bm \alpha$ directly determines the Monte Carlo sampling from the probabilistic model $P(\bm l,\bm \alpha)$. The gradient for the Monte Carlo expectations can be calculated by score function estimator. For normalized probability distributions, the gradient with respect to $\bm \alpha$ can be calculated by
\begin{eqnarray}
	\nabla_{\bm \alpha} \mathcal{L}=\sum_{\bm l \sim P}\nabla_{\bm \alpha} \ln P(\bm l,\bm \alpha) L(V(\bm l, \bm \theta)).
	\label{eq:gradientAlpha}
\end{eqnarray}
The continuous parameters $\bm \theta$ and $\bm \alpha$ are simultaneously optimized iteratively. After they converge,  a fine-tuning step on the gate parameters can achieve lower loss, \ie the optimal architecture $\bm l^*$ which has the highest probability in $P(\bm l,\bm \alpha)$ is fixed and the gate parameters are further optimized until convergence.

\section{Quantum architecture search with meta-learning}
\begin{figure}
	\centering
	\includegraphics[width=15cm]{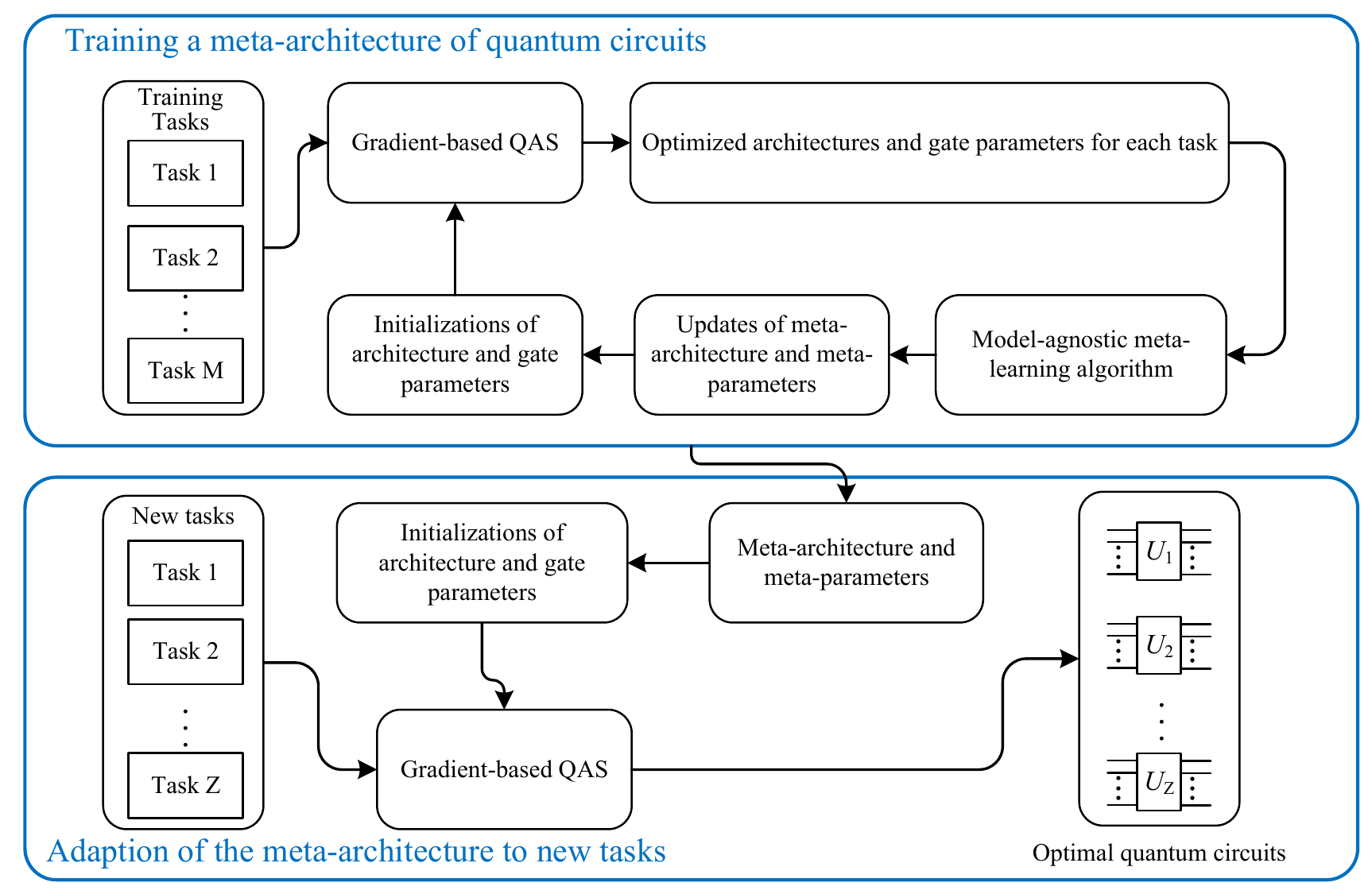}\\
	\caption{Overview of proposed meta quantum architecture search (MetaQAS) algorithm. MetaQAS consists of two parts:  (1) training of a meta-architecture and (2) adaption of the trained meta-architecture to new tasks. The meta-architecture is progressively updated according to the optimized architectures for a batch of training tasks. In each iteration, gradient-based QAS starts from the meta-architecture and outputs the optimized architectures for different training tasks. Then, the meta-architecture is updated based on these optimized architectures with a model-agnostic meta-learning algorithm.
		After the meta-training converges, the optimal meta-architecture which contains across-task knowledge is used to initialize the architecture of the gradient-based QAS algorithm for different new tasks and the optimal quantum circuits for these tasks can be generated by QAS with a small number of iterations.}
	\label{fig:MetaQAS}
\end{figure}

The proposed meta quantum architecture search (MetaQAS) algorithm divides into two steps: (1) training of a meta-architecture and (2) adaption of the trained meta-architecture to new tasks, as shown in \textbf{Figure \ref{fig:MetaQAS}}. In the training step, MetaQAS aims to learn a meta-architecture from a set of training tasks, which contains across-task knowledge.
The training tasks are drawn from a distribution over tasks $p(\mathcal{T})$ that we want our algorithm to be able to adapt to.
For quantum compiling, the training tasks include a set of target circuits.
In each training iteration, a mini-batch of tasks are randomly selected from the training tasks.
For each task, gradient-based QAS initializes the circuit architecture with the meta-architecture and outputs the optimized architecture.
Then the meta-architecture is updated according to the optimized architectures of different tasks.
The iteration repeats until convergence. In this way, MetaQAS progressively improves the adaptability by incorporating the training process of each task into the meta-architecture.
In the proposed MetaQAS, arbitrary gradient-based QAS and model-agnostic meta-learning algorithm can be used to learn the optimized architecture for each task and the meta-architecture. In this paper, we use DQAS\,\cite{zhang2020differentiable} and REPTILE\,\cite{nichol2018first}, respectively.
To further accelerate gradient-based QAS, we train the meta-parameters of quantum gates as well as the meta-architecture.
Once the meta-architecture is trained from a number of training tasks, it can be adapted to different new tasks.
For each new task, the gradient-based QAS algorithm starts from the meta-architecture and meta-parameters of quantum gates containing across-task knowledge instead of random initialization in current QAS algorithms, and can be fast adapted to a new task with just a few steps of gradient descent.

\subsection{Training of a meta-architecture}
In the training step, MetaQAS aims to find a meta-architecture $\bm \alpha_{meta}$ and meta-parameters $\bm \theta_{meta}$ of quantum gates, which satisfy
\begin{eqnarray}
	\nonumber (\bm \theta_{meta}, \bm \alpha_{meta}) &= \arg\min_{\bm \theta, \bm \alpha}\sum_{\mathcal{T}_i\sim p(\mathcal{T})}L_{\mathcal{T}_i}(U(\Phi^k(\bm \alpha,\bm \theta,\mathcal{T}_i)))\\
	&=\arg\min_{\bm \theta, \bm \alpha}\sum_{\mathcal{T}_i\sim p(\mathcal{T})}L_{\mathcal{T}_i}(U(\bm \alpha_{\mathcal{T}_i}^k, \bm \theta_{\mathcal{T}_i}^k)),
	\label{eq:meta}
\end{eqnarray}
where $L_{\mathcal{T}_i}$  is the loss function of task $\mathcal{T}_i$. MetaQAS can be applied to different applications by defining different loss functions $L$. For quantum compiling, it can be defined as the fidelity between the target circuit and the designed one.
$\Phi^k(\bm \alpha, \bm \theta, \mathcal{T}_i)$ denotes the learning algorithm at the $k$th iteration of updates from the initial parameters $(\bm \theta, \bm \alpha)$ for task $\mathcal{T}_i$.
The learning algorithm $\Phi^k$ not only optimizes the gate parameters $\bm \theta$, but also optimizes the architecture of the quantum circuit $\bm \alpha$ for a specific task.
$U(\bm \alpha_{\mathcal{T}_i}^k, \bm \theta_{\mathcal{T}_i}^k)$ denotes the quantum circuit determinated by the architecture $\bm \alpha_{\mathcal{T}_i}^k$ and the gate parameters $\bm \theta_{\mathcal{T}_i}^k$  at the $k$th iterations of updates for task $\mathcal{T}_i$.
As shown in Equation\,(\ref{eq:meta}), MetaQAS explicitly improves the adaptability by incorporating the training process of each task into the meta-objective.
By using a gradient-based QAS algorithm, the gate parameters and architecture can be updated with
\begin{eqnarray}
	\nonumber\bm \theta^{j+1} = \bm \theta^j - \lambda_{task} \triangledown_{\bm \theta} L_{\mathcal{T}_i}(U( \bm \alpha^j, \bm \theta^j)),\\
	\bm \alpha^{j+1} = \bm \alpha^j - \eta_{task} \triangledown_{\bm \alpha} L_{\mathcal{T}_i}(U( \bm \alpha^j, \bm \theta^j)),
	\label{eq:parTask}
\end{eqnarray}
where $U( \bm \alpha^j, \bm \theta^j)$ is the quantum circuit defined by the architecture $\bm \alpha^j$ and the gate parameters $\bm \theta^j$. $\lambda_{task}$ and $\eta_{task}$ are the learning rates of gate and architecture parameters for a given task.
In this paper, we use the state-of-the-art gradient-based QAS, namely DQAS \cite{zhang2020differentiable}, which embeds the discrete architecture choices into a continuously-parameterized probabilistic model. In DQAS, independent category probabilistic model is used, \ie
\begin{eqnarray}
	P(\bm l,\bm \alpha) = \Pi_i^N p(l_i, \bm \alpha_i),
	\label{eq:Pr}
\end{eqnarray}
where $\bm l=(l_1,l_2,...,l_N)$ denote the selected $N$ operations, which describe the selected $N$ quantum gates and the qubits they operate on. $\bm \alpha$ is of the dimension $N*N_o$, where $N_o$ is the number of operations which is determined by the numbers of candidate gates and the qubits they can act on.
The probability $p$ of selecting the $j$th operation in the $i$th layer is calculated by a softmax over all candidate operations in this layer
\begin{eqnarray}
	p(l_i=j, \bm \alpha_i) = e^{\alpha_{ij}}/\sum_k e^{\alpha_{ik}}.
	\label{eq:eq1}
\end{eqnarray}
We can sample a quantum circuit with $N$ gates according the probabilistic model $P(\bm l,\bm \alpha)$ in {Equation (\ref{eq:Pr})}. For DQAS, the gradients with respect to $\bm \theta$ and $\bm \alpha$ are calculated according to a batch of quantum circuits sampled from the probabilistic model $P(\bm l,\bm \alpha)$.

As gradient-based method is used to optimize the architecture and gate parameters for a task, we can use any gradient-based meta-learning algorithm, \eg MAML\,\cite{finn2017model} and REPTILE\,\cite{nichol2018first}, to solve the optimization problem in Equation\, (\ref{eq:meta}). In this paper, we use REPTILE due to its computational efficiency and satisfying performance. The meta gate and architecture parameters can be updated by
\begin{eqnarray}
	\nonumber  \bm \theta_{meta}^{j+1} = \bm \theta_{meta}^j + \lambda_{meta} \frac{1}{m}\sum_{i=1}^m(\bm \theta_{\mathcal{T}_i}^k - \bm \theta_{meta}^j ),\\
	\bm \alpha_{meta}^{j+1} = \bm \alpha_{meta}^j + \eta_{meta} \frac{1}{m} \sum_{i=1}^m(\bm \alpha_{\mathcal{T}_i}^k - \bm \alpha_{meta}^j),
	\label{eq:parMeta}
\end{eqnarray}
where $\lambda_{meta}$ and $\eta_{meta}$ are the learning rates of meta-parameters of quantum gates and architecture.
In each iteration, a mini-batch of $m$ tasks are randomly used to update the meta-architecture and meta-parameters of gates. \textbf{Algorithm\, \ref{alg:metalearning}} outlines the key steps of training a meta-architecture with DQAS and REPTILE. It should be noted that other gradient-based QAS algorithms and meta-learning algorithms can be used in the proposed algorithm instead of DQAS and REPTILE.
Eeach element in the architecture vector $\bm \alpha_{meta}$ is initialized with 0 and the gate parameters $\bm \theta_{meta}$ are initialized with random values range in [0,1].
In each meta-training epoch, a mini-batch of tasks are randomly drawn from the task training set $T$. The architecture $\bm \alpha_{\mathcal{T}_i}$ and gate parameters $\bm \theta_{\mathcal{T}_i}$ are optimized for each task $\mathcal{T}_i$ by using the meta-parameters $\bm \alpha_{meta}$ and $\bm \theta_{meta}$ as the starting points, and then are used to update the meta-architecture and meta-parameters of quantum gates.  The iteration repeats until convergence.
The meta-training process learns a meta-architecture and meta-parameters of quantum gates that are easy and fast to be fine-tuned, making the adaptation to new tasks starting at the right space for fast learning.
\begin{algorithm}
	\renewcommand{\algorithmicrequire}{\textbf{Input:}}
	\renewcommand\algorithmicensure {\textbf{Output:} }
	\caption{Training a meta-architecture with DQAS and REPTILE}
	\label{alg:metalearning}
	\begin{algorithmic}[1]
		\Require
		$T=\{\mathcal{T}_1,\mathcal{T}_2,...,\mathcal{T}_M\}$: training set with $M$ tasks;
		$\lambda_{task},\eta_{task}$: the learning rates of gate and architecture parameters for a given task;
		$\lambda_{meta},\eta_{meta}$: the learning rates of meta-parameters of quantum gate and architecture;
		\Ensure $\bm \alpha_{meta}, \bm \theta_{meta}$: the meta-architecture and meta-parameters of quantum gates.
		\State initialize $\bm \theta_{meta}$ with random values range in [0,1], initialize each element of $\bm \alpha_{meta}$ with 0.
		\Repeat
		\State randomly sample a mini-batch of tasks $T_b$ with $m$ circuits from the training set $T$.
		\For{each task $\mathcal{T}_i$ in $T_b$}
		\State $\bm \theta_{\mathcal{T}_i} \leftarrow \bm \theta_{meta}$
		\State $\bm \alpha_{\mathcal{T}_i} \leftarrow \bm \alpha_{meta}$
		\For {\texttt{$j=1$ to $k$}}
		\State update $\bm \theta_{\mathcal{T}_i}$ and  $\bm \alpha_{\mathcal{T}_i}$ according to Equation\,(\ref{eq:parTask}).
		\EndFor
		\EndFor
		\State update $\bm \theta_{meta}$ and  $\bm \alpha_{meta}$ according to Equation\,(\ref{eq:parMeta}).
		\Until{convergence}
	\end{algorithmic}
\end{algorithm}

\subsection{Adaption of the trained meta-architecture to new tasks}
The meta-training step has learnt a meta-architecture on different training tasks, which embeds the across-task knowledge. For unseen new tasks, QAS algorithm starts from the meta-architecture instead of random initialization on the architecture and only requires a small number of  updates for the quantum architecture before convergence. As model-agnostic meta-learning algorithm can learn initialization heuristics which can fast converge to local minimum of the objective function, MetaQAS can generate an optimal architecture for a new task much faster than state-of-the-art QAS algorithms.
\textbf{Algorithm\,\ref{alg:metalearningtest}} illustrates how it can be adapted to new tasks. For a new task, MetaQAS starts from the meta-architecture and meta-parameters of gates, and then the architecture and gate parameters are updated by gradient descent until convergence.  After that, the quantum circuit with the approximately optimal architecture can be generated by the highest probability in $P(\bm l, \bm \alpha_{\mathcal{T}})$.
As the architecture $\bm \alpha$ and the gate parameters $\bm \theta$ are optimized simultaneously, a fine-tuning step on the gate parameters will obtain a better quantum circuit.
Fine-tuning on the gate parameters means that we fix the optimal architecture $\bm l^*$ and further optimize the gate parameters by gradient descent until they converge.
\begin{algorithm}
	\renewcommand{\algorithmicrequire}{\textbf{Input:}}
	\renewcommand\algorithmicensure {\textbf{Output:} }
	\caption{Adaption of the meta-architecture to a new task}
	\label{alg:metalearningtest}
	\begin{algorithmic}[1]
		\Require
		$\mathcal{T}$: a new task;
		$\bm \alpha_{meta}$: the meta-architecture learned from a set of training tasks;
		$\bm \theta_{meta}$: the meta-parameters of quantum gates;
		$\lambda_{task},\eta_{task}$: the learning rates of gate and architecture parameters for a new task.
		\Ensure the optimal architecture and gate parameters for the new task $\mathcal{T}$.
		\State $\bm \theta_{\mathcal{T}} \leftarrow \bm \theta_{meta}$
		\State $\bm \alpha_{\mathcal{T}} \leftarrow \bm \alpha_{meta}$
		\Repeat
		\State $\bm \theta_{\mathcal{T}} \leftarrow \bm \theta_{\mathcal{T}} - \lambda_{task} \triangledown_{\bm \theta} L_{\mathcal{T}}(U(\bm \alpha_{\mathcal{T}}, \bm \theta_{\mathcal{T}}))$
		\State $\bm \alpha_{\mathcal{T}} \leftarrow \bm \alpha _{\mathcal{T}}- \eta_{task} \triangledown_{\bm \alpha} L_{\mathcal{T}}(U(\bm
		\alpha_{\mathcal{T}}, \bm \theta_{\mathcal{T}}))$
		\Until{convergence}
		\State get the optimal architecture $\bm l^*$ with the highest probability in $P(\bm l, \bm \alpha_{\mathcal{T}})$.
		\State fine-tune the gate parameters $\bm \theta_{\mathcal{T}}$ of the quantum circuit with  architecture $\bm l^*$ to get the optimal $\bm \theta_{\mathcal{T}}^*$.\\
		\Return the optimal circuit architecture $\bm l^*$ and its gate parameters $\bm \theta_{\mathcal{T}}^*$.
	\end{algorithmic}
\end{algorithm}

\section{Numerical simulation}
We explore the performance of MetaQAS on variational quantum compiling of three- and four-qubit circuits.
A meta-architecture and meta-parameters of quantum gates are first trained on the training task set according to \textbf{Algorithm\,\ref{alg:metalearning}}, and then are adapted to new tasks according to \textbf{Algorithm\,\ref{alg:metalearningtest}}.
The performance of MetaQAS is compared to DQAS algorithm\,\cite{zhang2020differentiable}, a state-of-the-art QAS algorithm, in terms of the number of required iterations and the loss function denoting the difference between the target circuit and the compiled one.

\subsection{Generation of target circuits for variational quantum compiling}
\label{sec:gen}
The target circuit $V_t$ is generated by randomly selecting $N$ gates from the gate set  $\mathcal{A}_{target} = \{R_X(3\pi/2), R_Y(3\pi/2), R_Z(3\pi/2), \text{S}, \text{T}, \text{CNOT},$ $ \text{CZ}, \text{CY}\}$ and the qubits they act on.
$R_X(3\pi/2), R_Y(3\pi/2), R_Z(3\pi/2), \text{S}$ and $\text{T}$ are single-qubit gates while $\text{CNOT}, \text{CZ}$ and $\text{CY}$ are two-qubit gates. These quantum gates can be operated on any qubit.
Considering the number of single-qubit gates in a quantum circuit is usually larger than that of two-qubit gates, we set the probabilities of selecting single- and two-qubit gates to be 0.7 and 0.3, respectively.
The number of operations in the gate set is $5n+3n(n-1)=3n^2+2n$, where $n$ is the number of qubits. For a target circuit with $N$ gates, the number of possible quantum circuits is $(3n^2+2n)^N$.
We consider three- and four-qubit target circuits with $4$ gates, resulting in $1,185,921$ and $9,834,496$ possible target circuits.
We randomly generate ten thousand target circuits to construct the training task set $T$, which is used to train the meta-architecture and meta-parameters of quantum gates.
To evaluate the performance of the proposed method, 100 target circuits are randomly generated for the new task set.

\subsection{Meta-training for variational quantum compiling}
\label{sec:metatraining}
\begin{figure}
	\centering
	\subfigure[Three-qubit circuits]{
		\includegraphics[width=9cm]{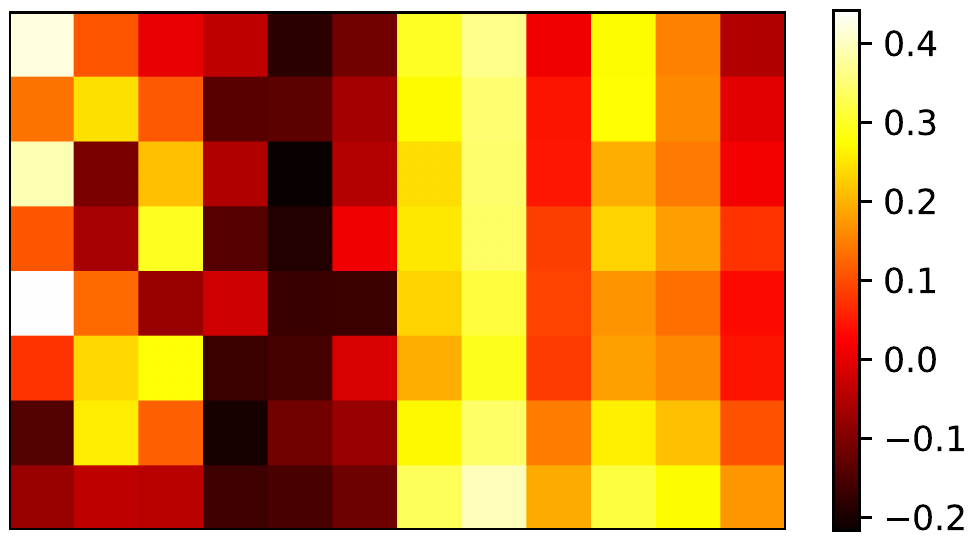}}
	\subfigure[Four-qubit circuits]{
		\includegraphics[width=12cm]{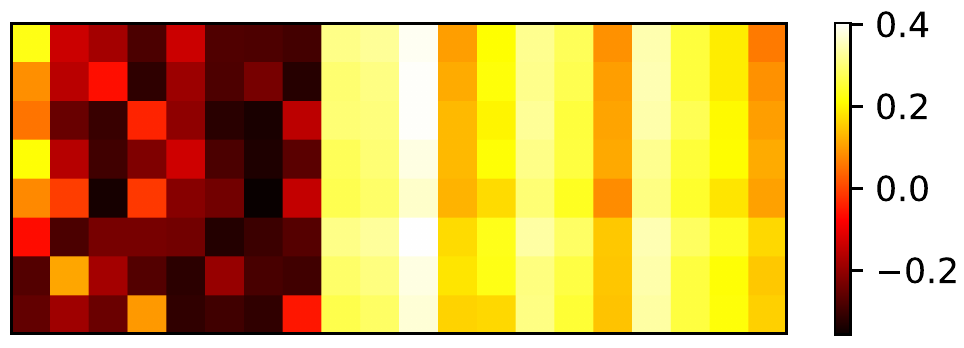}}
	\caption{The meta-architectures ($\bm \alpha_{meta}$) learnt by MetaQAS for three- and four-qubit circuits with 8 native gates. In DQAS, the architecture ($\bm \alpha$) is denoted by an $N$$*$$N_o$ matrix, where $N$ and $N_o$ are the number of gates in the quantum circuit and the number of candidate operations in each layer. The probability of selecting the $j$th operation in the $i$th layer can be calculated by $\alpha_{ij}$ according to Equation (\ref{eq:eq1}). A large $\alpha_{ij}$ indicates a high probability to select this operation.
		For three-qubit circuits, 1 to 3, 4 to 6, 7 to 12 columns denote the probabilities of selecting the quantum gates $R_X$, $R_Z$ and $\text{CNOT}$ gates acted on different qubits, respectively. For four-qubit circuits, 1 to 4, 5 to 8, 9 to 20 columns denote the probabilities of selecting the quantum gates $R_X$, $R_Z$ and $\text{CNOT}$ gates acted on different qubits, respectively.}
	\label{fig:metaArchitecture}
\end{figure}

The meta-architecture and meta-parameters of quantum gates are trained with the training task set $T$ according to \textbf{Algorithm\,\ref{alg:metalearning}}.
The native gate alphabet for quantum compiling is $\mathcal{A}=\{R_X(\theta), R_Z(\theta),\text{CNOT}\}$.
The target three- and four-qubit circuits are compiled with 8 native gates.
We use local Hilbert-Schmidt Test to measure the difference between the target and complied circuits.
The values of the hyperparameters during meta-training are shown in \textbf{Table\,\ref{Tab:hyperParameters}} in the Appendix.
In both three- and four-qubit circuits, the meta-architecture and meta-parameters of quantum gates at the $150$th iteration are used for the adaption to new target circuits.
The meta-architecture ($\bm \alpha_{meta}$) trained by MetaQAS for three- and four-qubit circuits with 8 native gates are shown in \textbf{Figure\,\ref{fig:metaArchitecture}}.
We can observe that most of  $RX$ and $RZ$ operations have low values in both three- and four-qubit circuits, which indicates that there are low probabilities to select rotation gates (\ie $RX$ and $RZ$) at the beginning. That is the reason that the loss of MetaQAS in the 0th iteration is much higher than that of DQAS as shown in \textbf{Figure\,\ref{fig:metaTrain}}. However, this kind of architectures may facilitate rapid convergence. This might be a useful heuristic initialization of architectures for gradient-based QAS algorithms.


\subsection{Adaption of the meta-architecture to new tasks}
\begin{figure}
	\centering
	\subfigure[Three-qubit circuits]{
		\includegraphics[width=8.9cm]{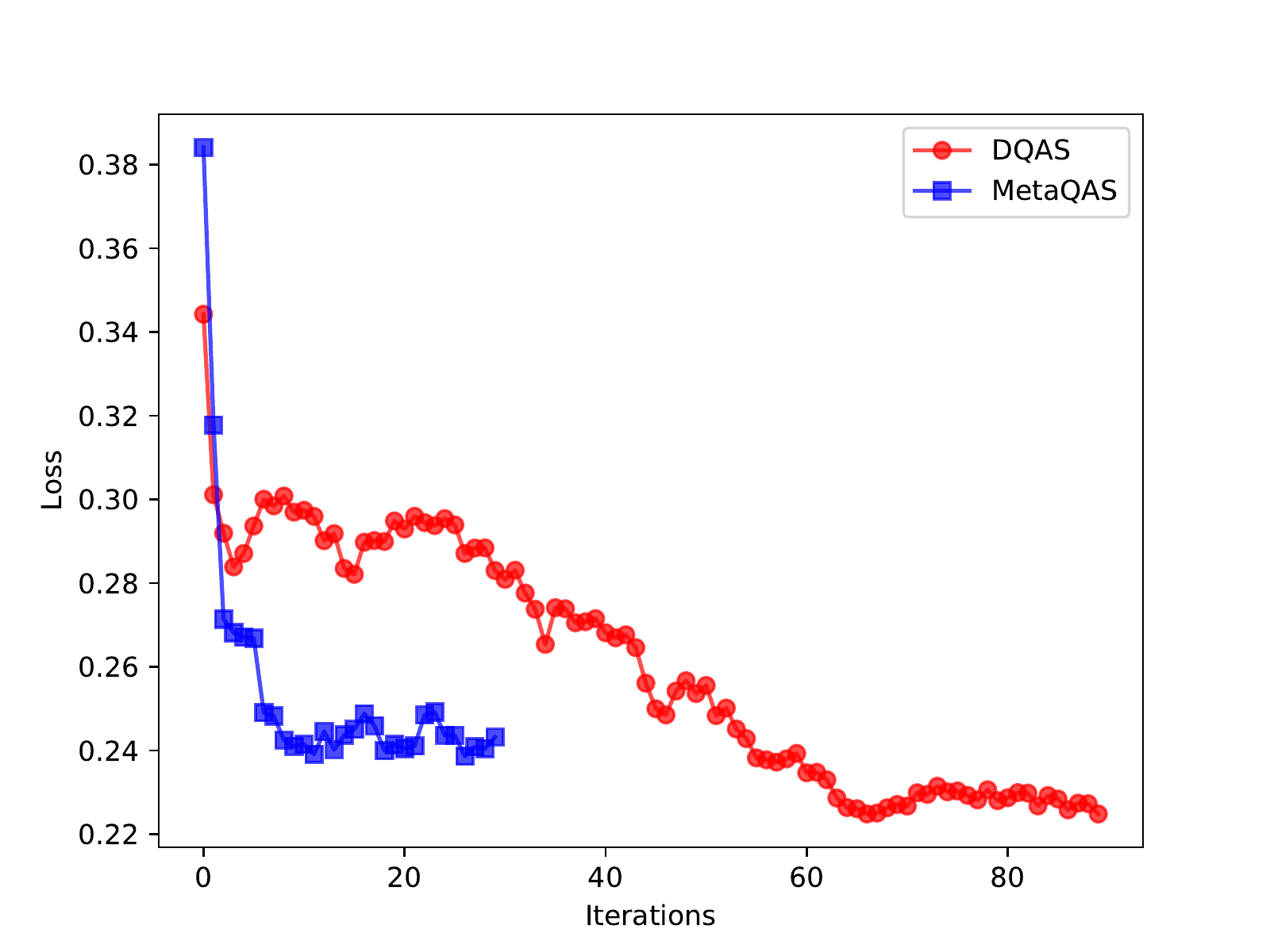}}
	\subfigure[Four-qubit circuits]{
		\includegraphics[width=8.9cm]{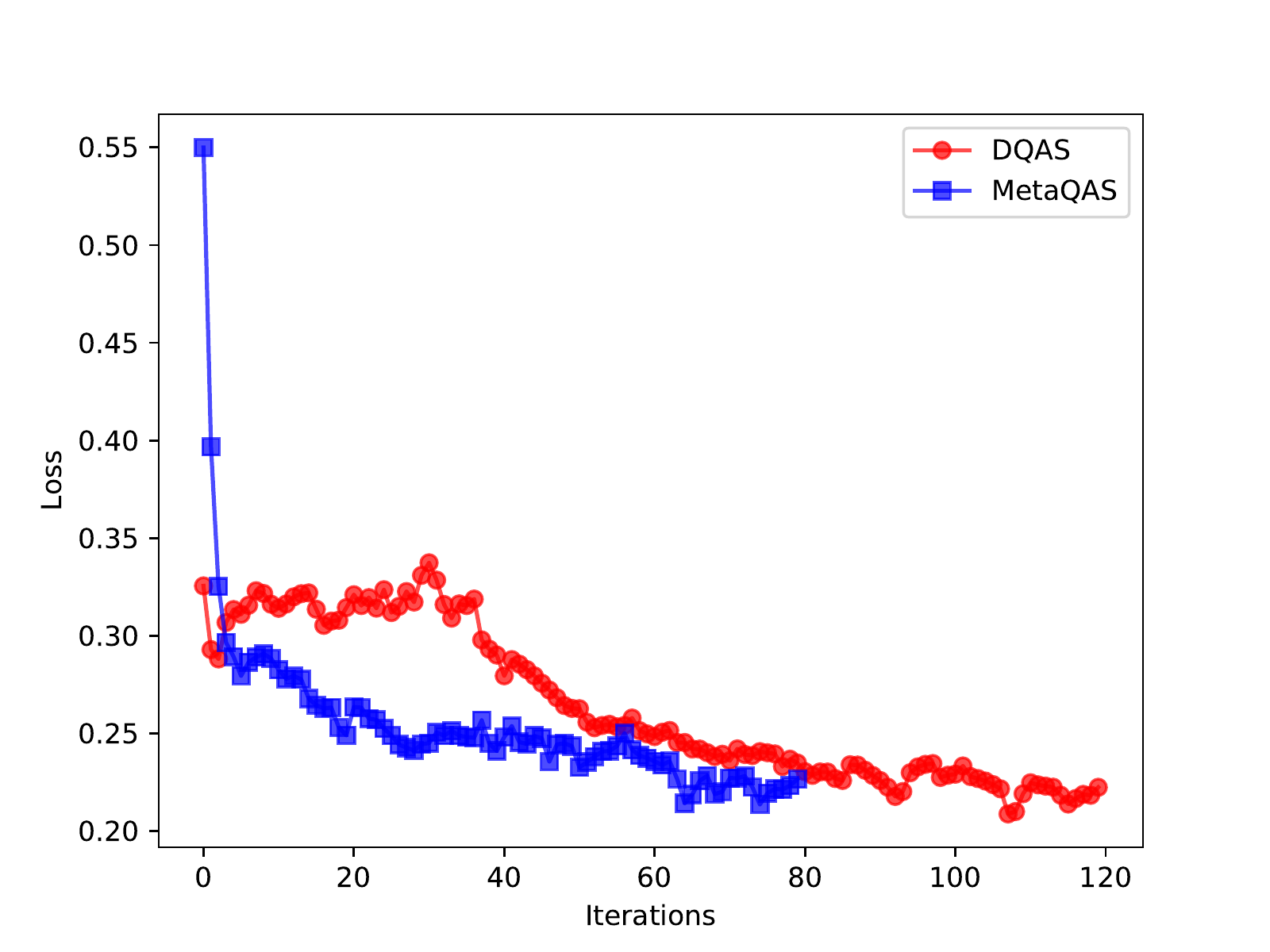}}
	\caption{The average loss of compiling 100 new target circuits by using different QAS algorithms with respect to the iterations of updates. In each iteration, we evaluate the loss using the quantum architecture $\bm l$ with the highest probability in the probabilistic model $P(\bm l, \bm \alpha_{\mathcal{T}})$. DQAS: the DQAS algorithm proposed in ref.\,\cite{zhang2020differentiable}; MetaQAS: the adaption algorithm of the meta-architecture to new tasks in \textbf{Algorithm \ref{alg:metalearningtest}}. In this figure, we only consider the quantum architecture search process without the fine-tuning step of the gate parameters.}
	\label{fig:metaTrain}
\end{figure}

In the task adaption step,  the trained meta-architecture and meta-parameters are used to initialize the architecture and gate parameters for DQAS algorithm and are adapted to new tasks according to \textbf{Algorithm\, \ref{alg:metalearningtest}}. The values of hyperparameters during adaption are shown in \textbf{Table\,\ref{Tab:hyperParameters}}.
The values of loss function in variational quantum compiling with respective to the iterations of updates are described in \textbf{Figure\,\ref{fig:metaTrain}}.
Local HST(\ie Equation\, (\ref{eq:lhst})) is served as the loss function as it is more suitable for circuits with large size than global HST\,\cite{KLP19}.
DQAS denotes the DQAS algorithm proposed in ref.\,\cite{zhang2020differentiable} while MetaQAS denotes the adaption algorithm of the meta-architecture to new tasks as described in \textbf{Algorithm \ref{alg:metalearningtest}}.
To make a fair comparison, both MetaQAS and DQAS use the same hyperparameters including batch size, training steps, learning rates of gate parameters and architecture.
The loss of MetaQAS drops much faster than that of DQAS with the increase of iterations and converges to a similar value in both three- and four-qubit circuits.
For the three-qubit case, the loss of MetaQAS converges after 30 iterations, only 1/3 of DQAS which needs 90 iterations.
The loss of MetaQAS in four-qubit case converges at the 80th iteration while the one of DQAS converges at the 120th iteration.
The results show that the meta-architecture containing across-task knowledge can enable rapid convergence to local minimum of the loss function.
MetaQAS has found a meta-architecture that is easy and fast to fine-tune, making the optimization of the quantum architecture for new tasks start at the right space for fast training and thus need fewer updates.

\begin{figure}
	\centering
	\subfigure[Three-qubit circuits]{
		\includegraphics[width=8.9cm]{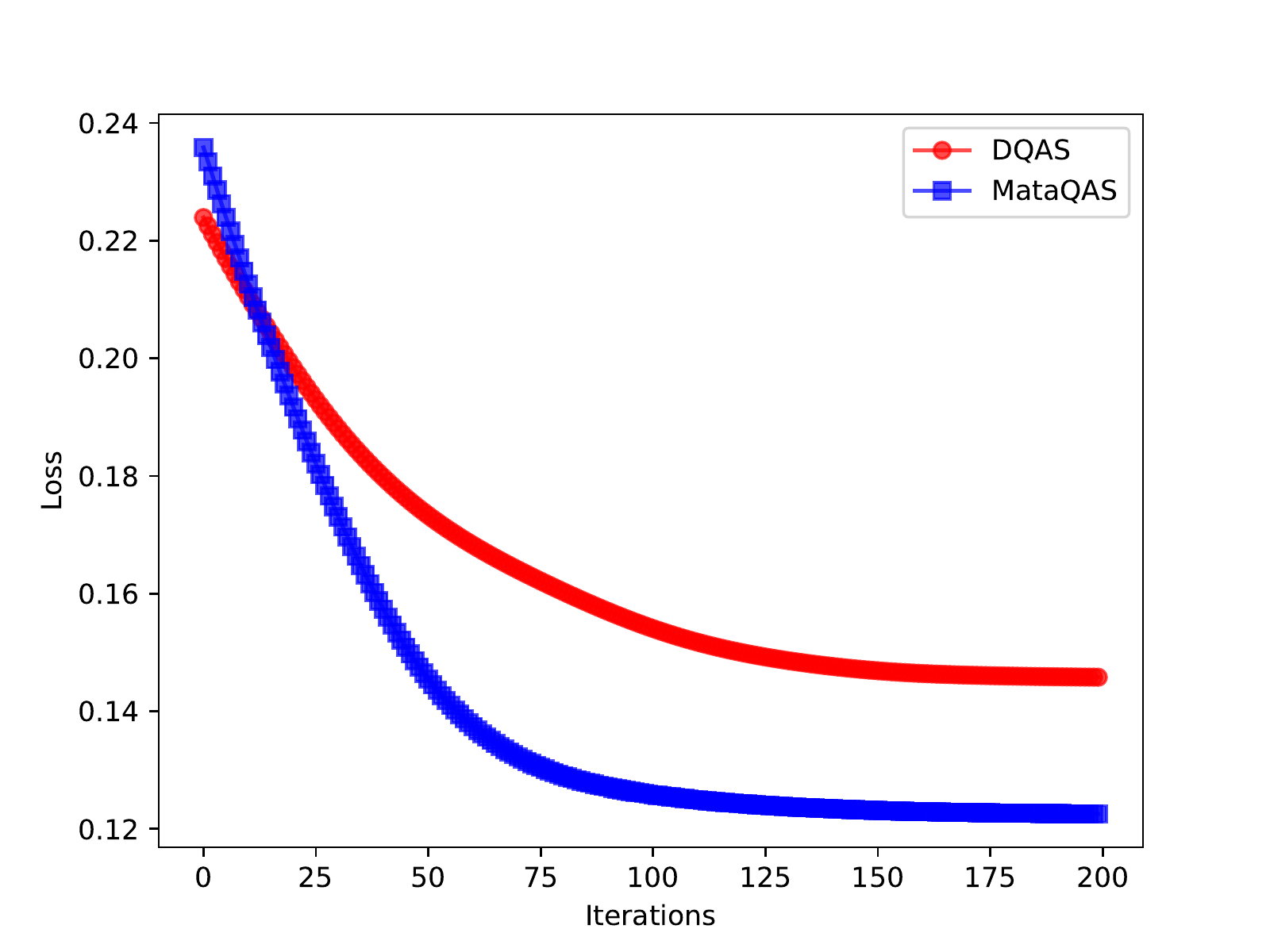}}
	\subfigure[Four-qubit circuits]{
		\includegraphics[width=8.9cm]{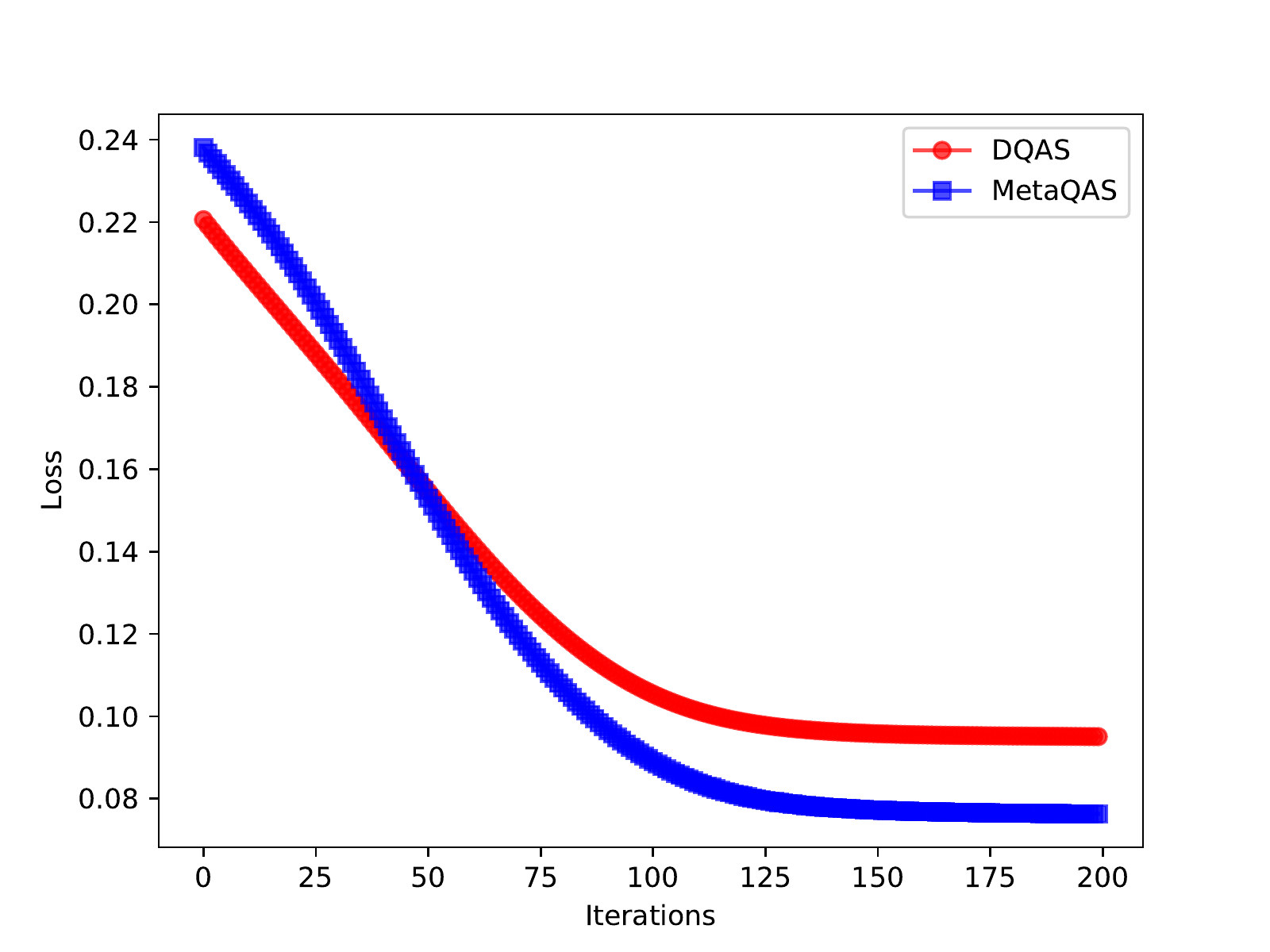}}
	\caption{The average loss of compiling 100 new target circuits by using different QAS algorithms with respect to the iterations of fine-tuning on the gate parameters.
		DQAS: the architecture and gate parameters in the $90$th ($120$th) iteration (\ie the last iteration in \textbf{Figure\,\ref{fig:metaTrain}}) of update by DQAS algorithm \cite{zhang2020differentiable} are used to be fine-tuned for three(four)-qubit circuits.
		MetaQAS: the architecture and gate parameters in the $30$th ($90$th) iteration of update by MetaQAS algorithm \cite{zhang2020differentiable} are used to be fine-tuned for three(four)-qubit circuits.}.
	\label{fig:metaFine}
\end{figure}

After the architecture searching process, a fine-tuning step is used to further optimize the gate parameters in both MetaQAS and DQAS algorithms.
We generate a quantum circuit with the highest probability in the probabilistic model $P(\bm l, \bm \alpha_{\mathcal{T}})$ as described in \textbf{Algorithm \ref{alg:metalearningtest}}.
The probabilistic models are the ones after 30(80) and 90(120) updates for MetaQAS and DQAS in the three-(four-)qubit quantum compiling, respectively.
The gate parameters of the generated quantum circuits after architecture searching process are further optimized and the losses with respect to the iterations are shown in \textbf{Figure\,\ref{fig:metaFine}}. We can observe that the fine-tuning step can further decrease the loss of VQC in both MetaQAS and DQAS. After fine-tuning step, the final loss of MetaQAS is lower than DQAS, which indicates that the trained meta-architecture and meta-parameters not only accelerate DQAS algorithm, but also enable DQAS algorithm to learn a better quantum circuit which gets lower loss. The reason might be that the meta-knowledge could make DQAS algorithm starting at the right space which avoids falling into local optimality.

We show the average loss of compiling 100 new target circuits after the fine-tuning step by DQAS and MetaQAS algorithms in \textbf{Table \ref{Tab:finetune}}.
In the simulation, the 100 target circuits are compiled with 8 native gates.
Both the average losses of VQCs based on DQAS and MetaQAS do not converge to 0 as some of the target circuits in the new tasks set are difficult to compile and may need more than 8 native gates to achieve zero loss.
\begin{table}
	\centering
	\caption{The average of loss functions after the fine-tuning step in DQAS and MetaQAS.  The 100 target circuits are random generated.  Although the difficulty of compiling these circuits varies and different numbers of native gates are required for compilation, we use a fixed number (\ie 8) of native gates to compile all the target circuits in the simulation.
		We use this setting as we focus on the improvement using the meta-architecture learnt by MetaQAS compared to the random initializations in DQAS in terms of convergence speed and the final loss. Both the average losses of VQCs based on DQAS and MetaQAS do not converge to 0 as some target circuits are complicated and need more than 8 native gates.}
	\begin{tabular}{ccc}
		\toprule
		&Three-qubit&Four-qubit\\
		\midrule
		DQAS&0.146&0.095\\
		MetaQAS&0.122&0.076\\
		\bottomrule
	\end{tabular}
	\label{Tab:finetune}
\end{table}

\begin{figure}
	\centering
	\subfigure[Three-qubit circuits]{
		\includegraphics[width=9.0cm]{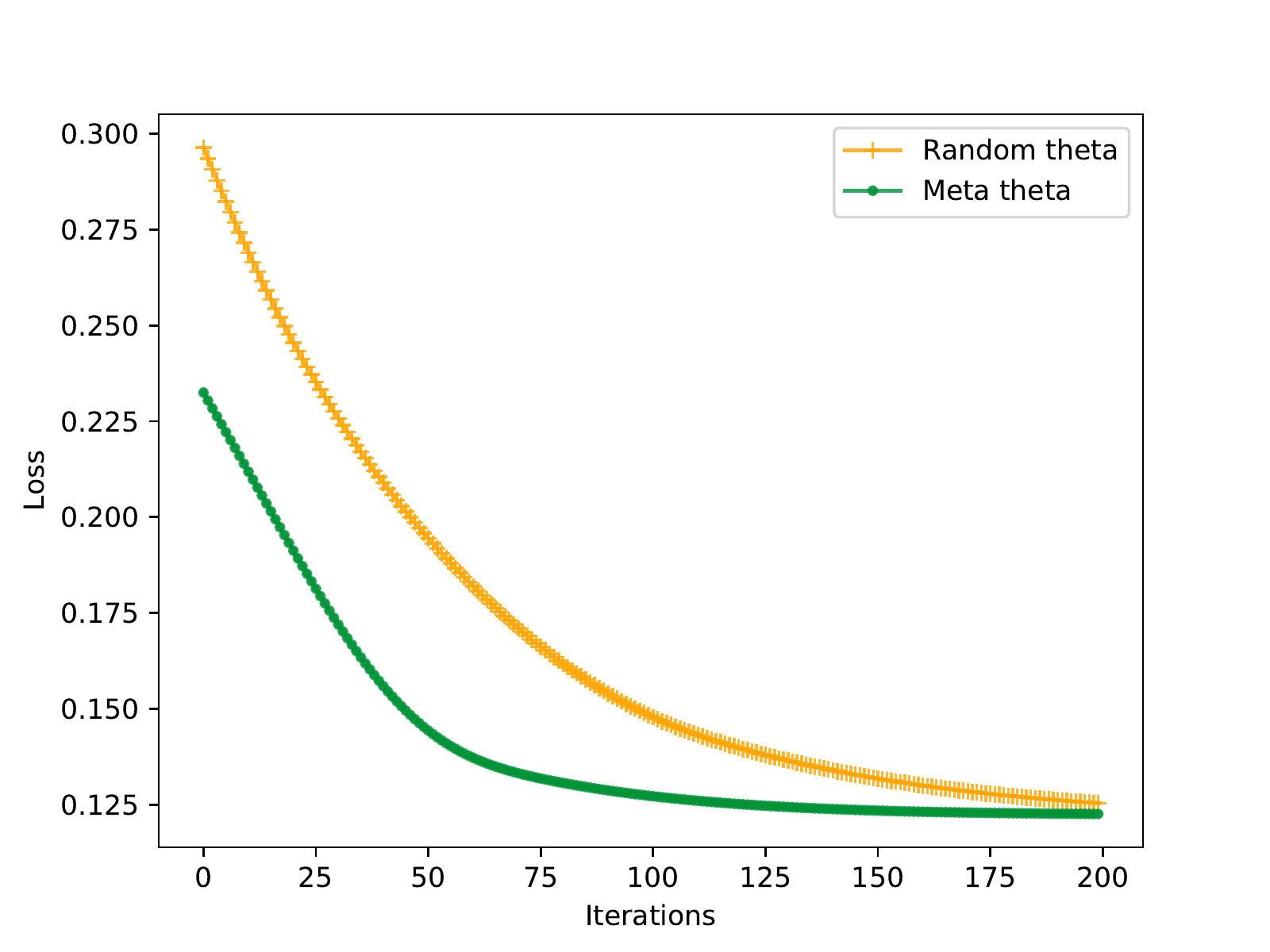}}
	\subfigure[Four-qubit circuits]{
		\includegraphics[width=9.0cm]{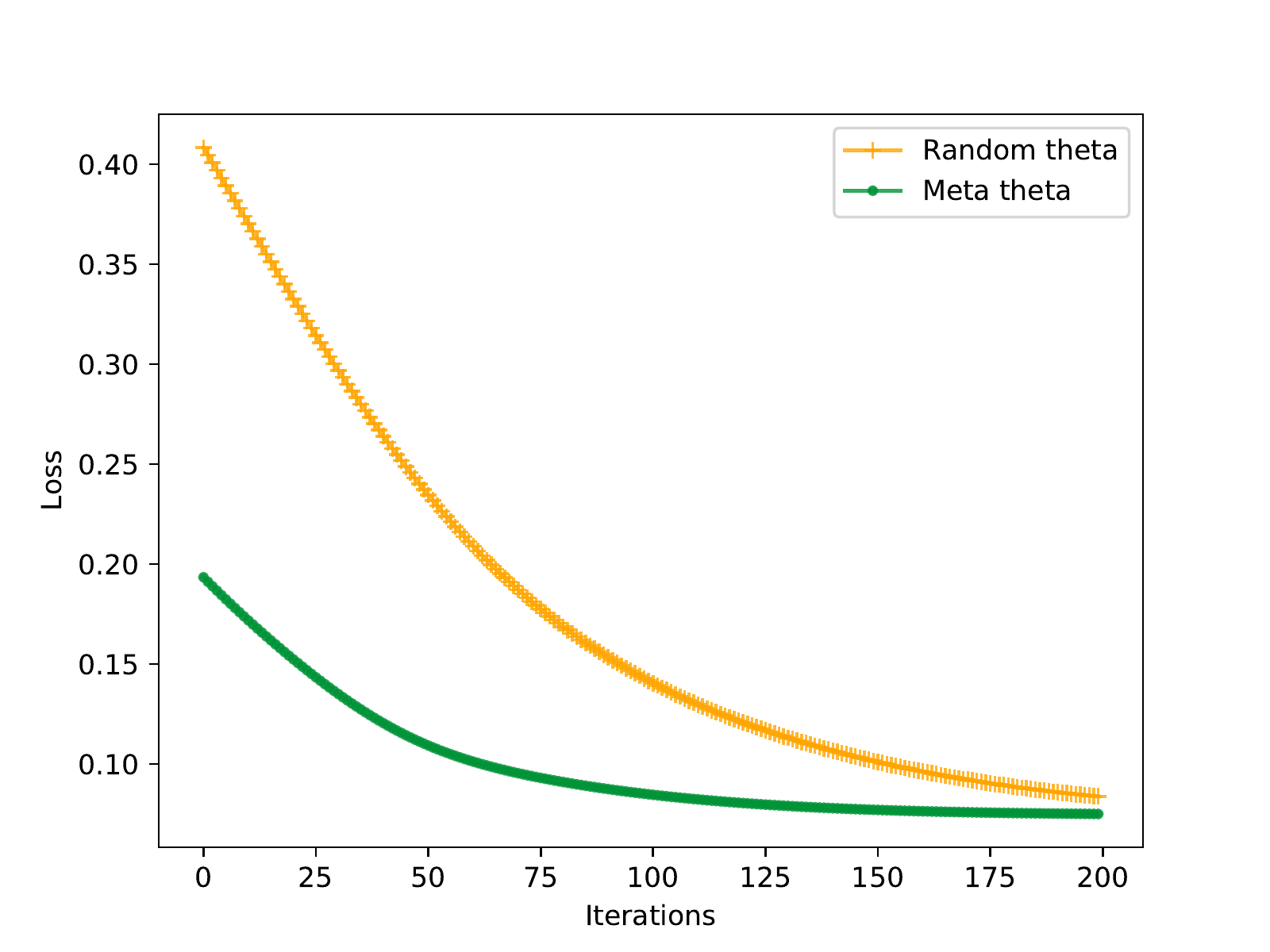}}
	\caption{The average loss of compiling 100 target circuits by using a fixed architecture with respect to the iterations of updating the gate parameters. Random theta: the quantum gates of the circuit are randomly initialized; Meta theta: the quantum gates are initialized by the meta gate parameters after meta-training as described in Section \ref{sec:metatraining}.}
	\label{fig:metaParameter}
\end{figure}

During the meta-training, meta-parameters of quantum gates are trained as well as the meta-architecture. We analyze if the meta-parameters of quantum gates can facilitate QAS algorithm.
We use the optimal quantum circuit after the architecture searching process of MetaQAS. Then the gate parameters of this circuit are initialized by different strategies, \ie random and meta strategies,  which denote that the gate parameters are initialized randomly and by the meta gate parameters after meta-training as described in Section \ref{sec:metatraining}, respectively. The averages of loss for random and meta strategies are illustrated in \textbf{Figure\,\ref{fig:metaParameter}}.
The loss of meta strategy decreases and converges much faster than the random one, especially for the four-qubit circuits, which indicates that the meta-parameters embedding across-task knowledge is easier to fine-tune.



\section{Conclusion}
In this paper, meta-learning is used to obtain across-task knowledge in order to find good initialization heuristics for the architecture and gate parameters of QAS algorithms.
The meta-architecture and meta-parameters of quantum gates are learned from a number of tasks, and can be fast adapted to new tasks with a small number of updates by gradient descent, which can speed up QAS algorithms.
The simulation results on variational quantum compiling of three- and four-qubit circuits show that the meta-architecture and meta gate parameters can not only accelerate DQAS algorithm, but also achieve a lower loss than DQAS.
The reason might be that MetaQAS could find a meta-architecture and meta-parameters that are easy and fast to fine-tune, making the optimization of the quantum architecture and gate parameters for a new task start at a right space for fast training and avoiding trapping into local optimum.

We will apply MetaQAS algorithm to other hybrid quantum-classical algorithms, \eg quantum approximate optimization algorithms and variational quantum machine learning in future work. In this paper, we only consider the gatewise QAS algorithm, which generates a quantum circuit gate by gate by selecting the gate type and position (\ie the qubit it acts on).
We will develop MetaQAS based on the layerwise QAS algorithm proposed in ref.\,\cite{zhang2021neural}, which generates a circuit by iteratively adding a layer of quantum gates.
The layerwise QAS algorithm can be directly transferred to larger systems. However, such direct transfer has no rigorous theoretical analysis and may suffer performance degradation. We will analyze how well the meta-architecture of layerwise QAS algorithm can transfer to larger systems.

\ack This work is supported by Guangdong Basic and Applied Basic Research Foundation (Nos.\,2021A1515012138, 2019A1515011166, 2020A1515011204, 2020B1515020050), the National Natural Science Foundation of China (Nos.\,61802061, 61772565, 61972091), the Cross Project of Foshan University (No.\,2019xw104), Key Platform, Research Project of Education Department of Guangdong Province (No.\,2020KTSCX132) and  the Key Research and Development Project of Guangdong Province (No.\,2018B030325001).
\section*{References}
\bibliographystyle{unsrt}

\clearpage
\appendix
\section{Hyperparameters of MetaQAS in the numerical simulation}
\label{sec:appendix}
\begin{table}[H]
	\centering
	\caption{Hyperparameters of MetaQAS in the simulations}
    \begin{tabular}{cccc}
  \hline \multirow{7}{*}{meta-training}
   &Hyperparameters &Three/Four qubits&Symbol\\

   \hline
   &learning rate of gate parameters for a task& 0.01 &$\lambda_{task}$\\
   &learning rate of architecture for a task& 0.1 & $\eta_{task}$\\
   &learning rate of meta gate parameters& 0.15 & $\lambda_{meta}$\\
   &learning rate of meta-architecture& 0.01 & $\eta_{meta}$\\
   &meta batch size & 100 & $m$\\
   &task training steps &5 & $k$\\
   &batch size in DQAS &256/512&$K$\\
   \hline \multirow{2}{*}{adaption}& learning rate of gate parameters for a task& 0.01/0.03&$\lambda_{task}$\\
   &learning rate of architecture for a task& 0.2/0.4& $\eta_{task}$\\
   &batch size in DQAS &256/512&$K$\\
    \hline
    \hline
    \end{tabular}
\label{Tab:hyperParameters}
\end{table}

\end{document}